\documentclass[aps, prb, twocolumn]{revtex4-2}

\usepackage[utf8]{inputenc}

\usepackage{ragged2e}
\usepackage{graphicx}
\usepackage{color}
\usepackage{dcolumn}
\usepackage{bm}
\usepackage{amsfonts}
\usepackage{amsmath}
\usepackage{amssymb}

\usepackage{url}
\usepackage{hyperref}
\usepackage{float}

\usepackage[dvipsnames]{xcolor}
\hypersetup{
	colorlinks,
	linkcolor={green!50!black},
	urlcolor={blue!80!black},
	citecolor   = {red!50!black} 
}

\usepackage{braket}
\usepackage{eucal}
\usepackage{color}
\usepackage{bm}
\usepackage{bbold}
\usepackage{subfig}

\def\be{\begin{equation}}
\def\ee{\end{equation}}
\def\bea{\begin{eqnarray}}
\def\eea{\end{eqnarray}}
\def\v{{\bf v}}
\def\ev{{\bf E}}
\def\bv{{\bf B}}

\def\kv{{\bf k}}
\def\q{{\bf q}}

\def\Sv{{\bf S}}

\def\Rv{{\bf R}}
\def\rv{{\bf r}}

\def\ve{\varepsilon}

\def\z{{\bf z}}
\def\s{{\boldsymbol\sigma}}

\usepackage{color}

\usepackage{soul}

\begin{document}


\title{Bilinear planar Hall effect in topological insulators due to spin-momentum locking inhomogeneity
}


\author{Amir~N.~Zarezad$^{1}$, J\'ozef~Barna\'s$^{1,2}$, Alireza~Qaiumzadeh$^{3}$, Anna~Dyrda\l{}$^{1 \ast}$} 
\affiliation{$^{1}$Faculty of Physics, ISQI, Adam Mickiewicz University in Pozna\'{n}, ul. Uniwersytetu Pozna\'{n}skiego 2, 61-614 Pozna\'{n}, Poland}
\affiliation{$^{2}$Institute of Molecular Physics, Polish Academy of Sciences, ul. M. Smoluchowskiego 17, 60-179 Pozna\'{n}, Poland}
\affiliation{$^{3}$Center for Quantum Spintronics, Department of Physics,  Norwegian University of Science and Technology,
NO-7491 Trondheim, Norway}
\email{adyrdal@amu.edu.pl}

\keywords{bilinear planar Hall effect, current-induced spin polarization, topological insulators, spin-momentum locking inhomogeneity
}

\begin{abstract}
We study the effect of spin-momentum locking inhomogeneity
on the planar Hall effect in topological insulators (TIs). Using the minimal model describing surface states of 3D TIs and semiclassical
Boltzmann formalism, we have derived the planar Hall conductivity within the generalized relaxation time approximation. We
have found that the total planar Hall conductivity is a sum of lin-
ear and nonlinear to the external electric field components. The linear term is a conventional planar Hall conductivity which scales quadratically with an external magnetic field, whereas the nonlinear term reveals bilinear behaviour, i.e., changes its sign when either
charge current density or in-plane magnetic field orientation is reversed. We have shown that the emergent nonlinear planar Hall effect is a consequence of spin-momentum locking inhomogeneity in the TIs with isotropic energy dispersion and dominates under the conventional planar Hall effect.
\end{abstract}

\maketitle

\section{Introduction}\label{sec:intro}
There is currently a great interest in topological phases  in condensed matter physics and material science communities, mainly because of their functionalities \cite{Gilbert2021,PhysRevLett.104.146802,PhysRevLett.100.096407,Ando13,Ando15,Masatoshi} and interesting physical phenomena \cite{Day2009ExoticST,MooreJoel,PhysRevLett.98.106803,PhysRevB.78.195424,PhysRevLett.102.146805,PhysRevLett.102.216403,PhysRevLett.102.216404,PhysRevLett.103.066402}. Topological insulators (TIs) are a class of topological materials with a strong intrinsic spin-orbit coupling (SOC) and an electronic bandgap in the bulk just as a normal insulator, while their surface states support emergent massless Dirac particles that are topologically  protected by time reversal symmetry \cite{Ando13,Ando15,RevModPhys.82.3045,RevModPhys.83.1057}. These surface electron states emerge inside the corresponding conventional insulating band gap, and their spin is locked perpendicularly to the relevant orbital momentum.
Spin-momentum locking makes TIs promising spintronic materials with a high efficiency of spin to charge conversion. Such a conversion appears due to the current-induced spin polarization \cite{EDELSTEIN1990233,Aronov1989,PhysRevLett.86.4358}, and is crucial for many phenomena, such as novel magnetotransport effects \cite{Wu2021} and spin–orbit torques (SOTs) \cite{PRB405,PRL601,Chernyshov,Mihai2010,PRB422,Woo2014}.

Recent studies on two-dimensional nonmagnetic materials with strong SOC have revealed a new type of nonlinear magnetoresistance in the presence of in-plane magnetic field or magnetization.
This 
nonlinear transport, called bilinear magnetoresistance (BMR), scales linearly
with both the applied electric field and magnetic  field \cite{Zhang_2018, He2018,PhysRevLett.124.046802,PhysRevMaterials.4.071001,BOBOSHKO2022168698,ZAREZAD2022169167,Boboshko2022}.

In the context of nonlinear magnetotranport effects observed for surface states of TIs,
S. Zhang and G. Vignale \cite{Zhang_2018, He2018} proposed a theoretical explanation of BMR as a consequence of nonlinear spin currents that apper in TIs with anisotropic Fermi contours, e.g. due to strong hexagonal warping like in surface states of $\mathrm{Bi_{2}Se_{3}}$. 
Another mechanism of BMR in TIs was proposed by  A.~Dyrda{\l} {\it et al.}~\cite{PhysRevLett.124.046802,BOBOSHKO2022168698,ZAREZAD2022169167,Boboshko2022}. According to this approach, BMR also appears in TIs with isotropic Fermi contours, and is a consequence of scattering on spin-orbital impurities or on structural defects (spin-momentum locking inhomogenities) in the presence of effective spin-orbital field due current-induced spin polarization.

In parallel to the nonlinear magnetoresistance, the nonlinear planar Hall effect has also been extensively studied recently in various systems with strong spin-orbit coupling.  P. He et al. have reported nonlinear planar Hall effect in $\mathrm{Bi_{2}Se_{3}}$, and interpreted their results by the mechanisms they proposed earlier for the  BMR effect (second-order non-linear spin currents in the presence of hexagonal warping) \cite{PhysRevLett.123.016801}. Later, it was shown that the nonlinear planar Hall effect can also result from electron-hole asymmetry in TIs under in-plane magnetic field as a direct consequence of the tilt and shift of Dirac cones~\cite{PhysRevB.103.155415,PhysRevB.101.041408}. Moreover, it has been shown recently that Zeeman-induced Berry curvature dipole may be another origin of nonlinear planar Hall effect~\cite{PhysRevResearch.3.L012006}.

In this Letter, we show that the mechanism proposed by A.~Dyrdal~et~al. for BMR in TIs
also leads to the  nonlinear planar Hall effect.
To do this, we 
restrict ourselves to the  minimal model describing surface states of TIs, i.e., we exclude  electron-hole asymmetry, hexagonal warping and hybridization between adjacent surface states in the sample. We show that the total planar Hall conductivity consists of a bilinear (nonlinear) term proportional to the current density $j$ and magnetic field $B$, and the term  proportional to $B^2$ that is a conventional planar Hall effect. 
Our numerical calculations show that the nonlinear term can be significantly larger than the linear one. To find the total planar Hall conductivity, we have employed the semiclassical Boltzmann formalism and used the generalized relaxation time approximation (GRTA) to solve it.
In Sec \ref{sec:model}, we present our model and introduce the relevant scattering processes. In Section \ref{sec:relaxation-time}, we explain in detail our approach and the method used to solve the Boltzmann equation. Then, in Sec 4 we describe the nonlinear planar Hall phenomenon.

\section{Model}
\label{sec:model}

We consider a single layer of a 3D TI, subject to an in-plane external magnetic field. 
To provide a possibly simple  picture of our mechanism, we neglect hybridization of the topological states at top and bottom surfaces of the layer, the hexagonal warping term, and the electron-hole asymmetry. 
In the presence of a charge current, a nonequilibrium spin polarization of charge carriers is induced in the system and we can write an effective Hamiltonian for a single surface in the form,
 \begin{equation}
\hat{H}^0_{\kv}=\hbar v_F(\kv\times\hat{\z})\cdot\s+\bv\cdot\s+\bv_{so}\cdot\s .
\label{eq:H0}
 \end{equation}
The first term in the Hamiltonian describes Dirac surface states, where $\kv$ is the 2D wave vector, $\hat{\z}$ denotes the unit vector normal to the surface of the TI layer, $\s=(\sigma_x,\sigma_y,\sigma_z)$ is the vector of Pauli matrices, $v_F$ is the Fermi velocity and $\hbar$ is the reduced Planck constant. The second term represents the  Zeeman energy in the external in-plane magnetic field,
$\bv=(B_x,B_y,0)= B (\cos\phi_B,\sin\phi_B, 0)$. $\bv$ is measured in energy unit, $\bv = g\mu_B {\bf b}$, where {\bf b}  is magnetic field, $g$ is the Lande g-factor, and $\mu_B$ is Bohr magneton.
The third term in the Hamiltonian, is a nonequilibrium Zeeman-like coupling between itinerant electrons and the effective spin-orbit field $\bv_{so}$ that
appears as a result of current-induced nonequilibrium spin polarization $\Sv$, $\bv_{so}=\mathcal{J}\Sv$, where $\mathcal{J}=-8\pi v_F/k_F$ and $k_F$ is the Fermi wave number~\cite{PhysRevLett.124.046802}.
The direction of the nonequilibrium spin polarization, and thus the effective spin-orbit field, is perpendicular to the average electron's momenta and is linear in the current density. The current-induced nonequilibrium spin polarization in a surface of TI is given by, $\Sv=-(\hbar^2/2e v_F)\, {\mathbf j} \times \hat{\z}$, where ${\mathbf j}$ is the in-plane charge current density and $e$ is the electron charge.

To explore the nonlinear planar Hall effect, we consider scattering processes arising from 
structural defects that introduce spin-orbital scattering processes and result in the spin-momentum locking
inhomogeneities, ${V}^{soc}(\rv)$, and extrinsic impurities described by scalar scattering potential, $V^{im}(\rv)$.
The scattering potential of spin-momentum locking inhomogenity 
can be written in the form
 \cite{ Winkler,PhysRevB.67.161303, PhysRevB.69.115333, PhysRevLett.104.256804, PhysRevB.93.081301},
\begin{equation}
V^{soc}(\rv)=-\frac{i}{2}\{\nabla_y,\alpha(\rv)\}\sigma_x+\frac{i}{2}\{\nabla_x,\alpha(\rv)\}\sigma_y,
\end{equation}
where $\alpha(\rv)$ describes fluctuation of the intrinsic spin-orbit coupling in the surface state of a TI. Correlations of these fluctuations are described by the white noise distribution, so we have $\langle\alpha(\rv)\rangle =0$ and $\langle\alpha(\rv)\alpha(\rv')\rangle =n_{s}\alpha^2\delta(\rv-\rv')$, where $\langle\dots\rangle$ denotes ensemble average, $n_{s}$ is the concentration of scattering centers, and $\alpha$ is treated as a free parameter in our formalism.
In turn, scattering by spin-independent scalar impurities 
in our system is described by 
\begin{eqnarray}
V^{\rm{im}}(\rv)=
 \sum_{i} V_0^i\sigma_0
 \delta(\rv-\Rv_i),
\label{Vim}
\end{eqnarray}
where $V_0^i$ is scattering potential  of the $i$th impurity located at the position  $\Rv_i$.
We consider white noise correlations for the impurities, so $\langle V^{\rm{im}}(\rv)V^{\rm{im}}(\rv')\rangle=n_{i}v_0^2\delta(\rv-\rv')$ 
where $n_i$ is the concentration of impurities.

In the Bloch states, the total scattering potential can be written as
$V^{\mathrm{scatt}}_{\mathbf{k} \mathbf{k}'} = V^{\rm{soc}}_{\mathbf{k} \mathbf{k}'} + V^{\rm{im}}_{\mathbf{k} \mathbf{k}'}$,
where
\begin{align}
&V_{\mathbf{k}\mathbf{k}'}^{\rm{soc}} = \frac{\alpha}{2} \left[ (k_{y} + k'_{y}) \sigma_{x} - (k_{x} + k'_{x}) \sigma_{y}\right], \label{Vsoc} \\
&V^{\rm{im}}_{\mathbf{k} \mathbf{k}'} = v_{0} \sigma_{0} .
\end{align}

We consider a charge current along the $x$ direction, so $\bv_{so}=(0, B_{so}, 0)$. Then,
upon a gauge transformation, $k_y=q_y-\frac{{B}_x}{v_F}, k_x=q_x+\frac{{B}_y}{v_F}+\frac{B_{so}}{v_F}$, the  Hamiltonian (\ref{eq:H0}) and the scattering potential due to the SOC inhomogeneity (\ref{Vsoc}), respectively, become,
 \begin{align}
\hat{H}^0_{\q}&=\hbar v_F(\q\times\hat{\z})\cdot\s,
\label{eq:H0q}\\
 V^{soc}_{\q\q'}&=\frac{\alpha}{2}[(q_y+q'_y)\sigma_x-(q_x+q'_x)\sigma_y]-\frac{\alpha}{v_F}{\tilde{\bv}}\cdot\s
 \label{v_soc_gauge}
 \end{align}
where $\tilde{\bv}=\bv+\bv_{so}$. Accordingly, after gauge transformation the scattering potential due to  spin-momentum locking inhomogeneity depends on the effective magnetic field in the system. In turn, scattering by the potential of impurities (\ref{Vim}) remains unchanged.
The main  objective of our work is to study the influence of scattering processes 
on the  planar Hall effect. For this purpose, we use the semiclassical Boltzmann approach, as described in the following section.

\section{Semiclassical Boltzmann approach}
\label{sec:relaxation-time}

The nonequilibrium Fermi-Dirac distribution function $f$  in the presence of an applied electric field $\ev$  can be found via the following steady state Boltzmann equation 
in the form \cite{PhysRevB.68.165311}
\begin{equation}
e\v_{\q}\cdot\ev\left(\frac{\partial
f^0}{\partial\ve}\right)=\left(\frac{\partial f}{\partial t}\right)_{\rm{coll}},\label{eq:boltzmann1}
\end{equation}
where ${\bf v}_\q=v_\q(\cos\phi,\sin\phi)$ is the group velocity of Dirac electrons, and $f^0$ and $f$ are the equilibrium and nonequilibrium Fermi-Dirac distribution functions, respectively. Considering elastic scattering and
detailed balance condition, one can write the collision term $(\partial f/\partial t)_{\rm{coll}}$ in the following form,
\begin{equation}
(\frac{\partial f}{\partial t})_{\rm{coll}}=
\int
\frac{d^2 \q'}{( 2\pi/L)^{2}}w(\q,\q')[f(\q,{\bf E})-f(\q', {\bf
E})], \label{eq:boltzmann2}
\end{equation}
where $L^2$ is the system area, while $w(\q,\q')$ is the transition rate between two eigenstates
($\ket{\q}$ and $\ket{\q'}$) of the Hamiltonian $\hat{H}^0_{\q}$. Using
Fermi golden rule, the transition rate can be written in terms of the
scattering amplitude $|T_{\q,\q'}|^2$ as follows,
\begin{equation}
w(\q,\q')=\frac{2\pi}{\hbar}|T_{\q,\q'}|^2\delta(\ve_q-\ve_{q'}).
\end{equation}

Up to the first Born approximation, the $T$-matrix is given by,
$T_{\q,\q'}\approx\langle\q|V^{\rm{scatt}}|\q'\rangle$. Since there is no correlation between scattering sources in the system, thus $\langle|T_{\q,\q'}|^2\rangle =\langle|T^{\rm{soc}}_{\q,\q'}|^2\rangle  +\langle|T^{\rm{im}}_{\q,\q'}|^2\rangle $. Finally, the total transition rate is given by the sum of the transition rates due to the separate mechanisms according to the Matthiessen's rule \cite{Ashcroft}:
\begin{equation}
w(\q,\q')=w^{\rm{soc}}(\q,\q')+w^{\rm{im}}(\q,\q'),
\label{Wtot}
\end{equation}
with,
\begin{eqnarray}
w^{\rm{soc}}(\q,\q')=\frac{2\pi}{\hbar}|T^{soc}_{\q,\q'}|^2\delta(\ve_q-\ve_{q'}),\\
w^{\rm{im}}(\q,\q')=\frac{2\pi}{\hbar}|T^{im}_{\q,\q'}|^2\delta(\ve_q-\ve_{q'}).
\end{eqnarray}

Combining Eqs. (\ref{eq:boltzmann1}) and (\ref{eq:boltzmann2}), we arrive at the following Boltzmann equation,
\begin{equation}
e\v_{\q}\cdot\ev\left(\frac{\partial f^0}{\partial\ve}\right)=\int
\frac{d^2\q'}{( 2\pi/L)^{2}}w(\q,\q')[f(\q,{\bf E})-f(\q', {\bf E})]. \label{eq:boltzmann3}
\end{equation}
There are several well-known methods proposed to solve the Boltzmann equation for anisotropic systems~\cite{PhysRevB.79.045427,PhysRevB.68.165311,PhysRevLett.99.147207,PhysRevB.75.155323}. Here we apply the generalized relaxation time approximation (GRTA), as implemented recently by V\'yborn\'y et al.~\cite{PhysRevB.79.045427}. In this approach  the nonequilibrium distribution function is expanded as follows;
\begin{equation}
f-f^0=e E v_\q\left(\frac{\partial f^0}{\partial
\varepsilon}\right)[a(\phi)\cos\chi+ b(\phi)\sin\chi],\label{eq:relax-time-anisotropic}
\end{equation}
where $\chi$ is the polar angle of the electric field and $a(\phi)$ and $b(\phi)$ have the meaning of relaxation times along $x$ and $y$ axis, respectively.
Inserting Eq. (\ref{eq:relax-time-anisotropic}) into Eq. (\ref{eq:boltzmann3}), one finds the system of  two decoupled inhomogeneous Fredholm equations,
\begin{subequations}\label{eq:sinphi}
\begin{align}
&\cos\phi=\bar{w}(\phi)a(\phi)-\int d\phi' w(\phi,\phi')a(\phi'), \\
&\sin\phi=\bar{w}(\phi)b(\phi)-\int d\phi' w(\phi,\phi')b(\phi'), 
\end{align}
\end{subequations}
where $\bar{w}(\phi)=\int d\phi' w(\phi,\phi')$ and
$w(\phi,\phi')=\frac{1}{(2\pi/L)^2}\int q'dq'w(\q,\q')$. Solving the Fredholm equations (\ref{eq:sinphi}), we find $a(\phi)$ and $b(\phi)$, and consequently the total distribution function.

\section{Nonlinear planar Hall effects}\label{sec:planer Hall effects}

The conductivity tensor components are given by,
\begin{equation}
\sigma_{ij}=\frac{e}{E_j}\int \frac{d^2k}{(2\pi)^2}v^{i}_{\q} f(\q,\ev),
\label{sigmaij}
\end{equation}
where  $i$ ($j$) refers to the response current (applied electric field) direction. Here, we are interested in the 
transverse current response of the system to an applied electric field. Using the nonequilibrium distribution function, Eq. (\ref{eq:relax-time-anisotropic}), we find the following relation for the transverse conductivity 
\begin{equation}
\sigma_{yx}=\frac{e^2}{(2\pi)^2}\int v_\q^2 a(\phi)\sin\phi\left(\frac{\partial
f^0}{\partial \ve}\right) q\, dq\, d\phi.
\label{eq:sigma-yy}
\end{equation}
In general, one can decompose this transverse conductivity 
as the sum of a symmetric and an asymmetric parts, $\sigma_{yx}=\sigma^{as}_{yx}+\sigma^{s}_{yx}$,
where we define the asymmetric conductivity as $\sigma^{as}_{yx}=[\sigma_{yx}(j_x)-\sigma_{yx}(-j_x)]/2$, and the symmetric conductivity as $\sigma^{s}_{yx}=[\sigma_{yx}(j_x)+\sigma_{yx}(-j_x)]/2$. Up to first order in the applied electric field, the symmetric part is independent of electric field and quadratic in magnetic field, which is the conventional planar Hall effect in linear response regime. The asymmetric part is the nonlinear response to the external electric field, and reveals bilinear behaviour, i.e., it scales simultaneously linearly with applied magnetic and electric fields. 
At low temperatures, we have $\partial f^0/\partial\ve \approx \delta(\ve-\ve_{\rm F})$, and we find,
\vspace{-0.1cm}
\begin{equation}
\sigma_{yx}=\frac{v_Fk_F}{4\pi }c_2 \sigma_0 =\sigma_{0}\frac{\varepsilon_{F}}{2h} c_{2},
\label{eq:sigma-yx}
\end{equation}
where $c_2=\int d\phi \big(a(k_F,\phi)\sin\phi\big)$ is a numerical coefficient, found from the Fredholm equations (\ref{eq:sinphi}), $\varepsilon_F$ is the Fermi energy, and $\sigma_0=\frac{2 e^2}{h}$ is the conductance quantum.
\begin{figure}[t]
      \centering
      \includegraphics[width=0.90\linewidth]{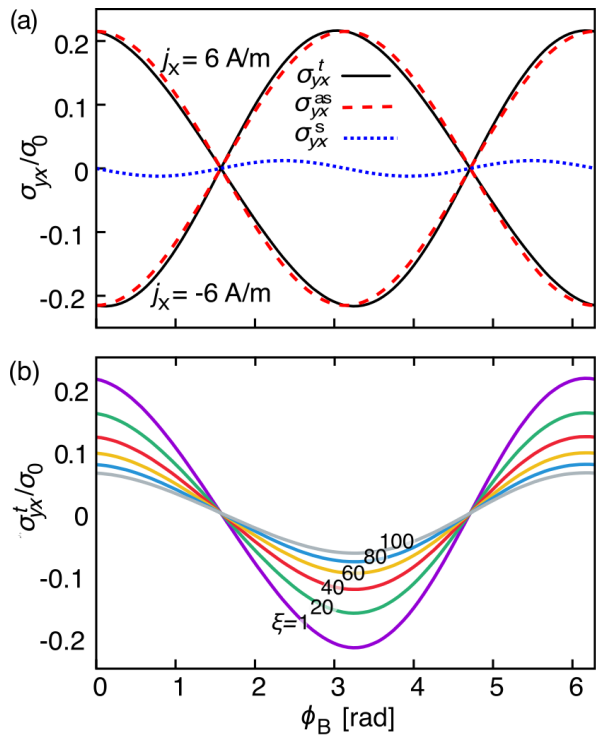}
      \caption{(a) Total $\sigma_{yx}$ (solid lines) and the corresponding nonlinear (dashed lines ) and linear (dotted line) contributions, presented versus $\phi_B$  for two opposite current densities and $\xi=1$ (here $\xi = n_iv_0^2 / n_iV_0^{2({\rm ref})}$, with $n_iV_0^{2({\rm ref})}$ being a reference value). (b) Total conductivity $\sigma_{yx}$ vs $\phi_B$ in the absence of  magnetic impurities for different values of $\xi$. Here, b=2 T, $j_x=6 A/m$, $\varepsilon_F=50 meV$. }
      \end{figure}
Upon  expansion up to the linear terms in the current density and quadratic terms in the magnetic field, the total transverse conductivity is given by the following simple analytical expressions:
\begin{subequations}
\begin{align}
&\sigma^{t}_{yx}=\sigma^{as}_{yx}+\sigma^{s}_{yx}, \\
&\sigma^{as}_{yx}=\dfrac{2n_s\alpha^2 /\sigma_{0}}{(n_s\alpha^2k_F^2+n_iv_0^2)^2}\frac{4\pi \hbar^2  v_F}{e\varepsilon_F}j_x B\cos\phi_B,\\
&\sigma^{s}_{yx}=-\dfrac{n_s\alpha^2 /\sigma_{0}}{(n_s\alpha^2k_F^2+n_iv_0^2)^2} B^2\sin2\phi_B,
\label{conductivity-V0}
\end{align}
\end{subequations}
where $j_x$ is the applied charge current density along the $x$ axis. The prefactors on the right sides clearly show that the planar Hall effect is generated by spin-orbit scattering as the Hall conductivity (both symmetric and asymmetric components) vanishes in the limit of $n_s\alpha^2\to 0$. Moreover, these prefactors also show that scattering on scalar impurities reduces the planar Hall conductivity.  From the above equations also follows that the bilinear (asymmetric) part, proportional to $\bm{B}\cdot \bm{j}=B j_x \cos\phi_B$, is enhanced at small Fermi energies as it is inversely proportional to $\epsilon_{F}$ (disregarding the impact of $\varepsilon_{F}$ in the prefactor describing scattering). 
The symmetric term, is the conventional linear planar Hall effect, which is quadratic in the applied magnetic field, $B_x B_y \sim B^2 \sin{2\phi_B}$.
The periodicity of the nonlinear term with respect to the magnetic field direction $\phi_B$ is $2\pi$, while the periodicity of the linear planar Hall effect is $\pi$.

The corresponding numerical results are shown in Fig. 1. The total planar Hall conductivity, Eq. (\ref{eq:sigma-yx}), is shown by the solid lines for two opposite directions of the applied current density. The corresponding contributions from the nonlinear and linear terms of the planar Hall conductivity are presented by the dashed and dotted lines, respectively. For the parameters assumed there, the nonlinear term is dominating over the conventional linear planar Hall effect. Interestingly,the two different periodicities are clearly visible in this figure. Experimentally, the nonlinear term can be extracted as an asymmetric part of the measured planar Hall conductivity.

The total planar Hall conductivity is reduced with increasing the spin-independent electron scattering rate. This dependence is presented explicitly in Fig. 1(b), where the total planar Hall conductivity is shown as function of polar angle of the in-plane magnetic field for different values of the nonmagnetic impurity scattering potential, that is quantified by the parameter $\xi$.

Behavior of the  bilinear planar Hall conductivity with the magnetic field and current density, and its dependence on the impurity scattering rate, is similar to that for BMR, associated with diagonal components of the conductivity tensor. This is because physical origins of the nonlinear planar Hall effect are the same as those for the BMR effect.

\section{Summary and Conclusions}

We have analysed the planar Hall effect in 3D TIs, where the surface topological electron states are described by the  minimal model. We proposed a physical mechanism of the nonlinear planar Hall conductivity, which is based on the electron scattering by spin-momentum locking  inhomogeneitiy in the presence of an effective spin-orbital field that appears due to nonequilibrium spin polarization induced by an external electric field. 
We have identified  two terms in the planar Hall conductivity: the linear and nonlinear ones with respect to the external electric field. The nonlinear term is proportional to the applied current density and external magnetic field (bilinear term), and reveals the $2\pi$-periodicity with respect to the orientation of in-plane magnetic field. The linear term has the well known quadratic dependence on the magnetic field and shows the $\pi$-periodicity with respect to the orientation of the in-plane magnetic field. We also showed that the spin-momentum locking inhomogeneity is crucial in our theory for achieving a nonzero planar Hall conductivity, whereas the scalar impurities reduce the planar Hall effect.

\medskip
\medskip
\textbf{Acknowledgements} \par 
This work has been supported by the Norwegian Financial Mechanism under the Polish-Norwegian Research Project NCN GRIEG '2Dtronics', project no. 2019/34/H/ST3/00515. A. Q. was partially supported by the Research Council of Norway
through its Centres of Excellence funding scheme, Project No. 262633, 'QuSpin'.

\medskip
{\small{

}}
\end{document}